\documentclass[conference]{IEEEtran}
\IEEEoverridecommandlockouts
\usepackage{cite}
\usepackage{caption}
\usepackage[none]{hyphenat}
\usepackage{algpseudocode}
\usepackage{amsmath}
\usepackage{amsfonts}
\usepackage{amssymb}
\usepackage{subfigure}

\usepackage{graphicx}
\usepackage[table]{xcolor}
\usepackage{array,multirow}

\usepackage{multicol}

\usepackage{amsfonts}
\usepackage{booktabs}
\usepackage{siunitx}

\usepackage{color}

\usepackage{tikz}

\usepackage{enumitem}

\usepackage{url}
\makeatletter
\def\url@leostyle{%
  \@ifundefined{selectfont}{\def\UrlFont{\sf}}{\def\UrlFont{\small\ttfamily}}}
\makeatother
\newcommand{\eat}[1]{}
\usepackage[linesnumbered,ruled]{algorithm2e}
\usepackage{mdframed} 

\renewcommand{\IEEEbibitemsep}{1pt plus 2pt}
\makeatletter
\IEEEtriggercmd{\reset@font\normalfont\small}
\makeatother
\IEEEtriggeratref{1}

\usepackage{xcolor}
\definecolor{light-gray}{gray}{0.9}

\usepackage{amsthm}

\def\BibTeX{{\rm B\kern-.05em{\sc i\kern-.025em b}\kern-.08em
    T\kern-.1667em\lower.7ex\hbox{E}\kern-.125emX}}

\begin{document}

\title{Measuring Decentralization in Bitcoin and Ethereum using Multiple Metrics and Granularities
}

\author{
  \IEEEauthorblockN{
    Qinwei~Lin\IEEEauthorrefmark{1},
    Chao~Li\IEEEauthorrefmark{1}\IEEEauthorrefmark{2},
    Xifeng~Zhao\IEEEauthorrefmark{1} and
    Xianhai~Chen\IEEEauthorrefmark{1}
  }
  \IEEEauthorblockA{
    \IEEEauthorrefmark{1}School of Computer and Information Technology, Beijing Jiaotong University, China \\
    \IEEEauthorrefmark{2}Beijing Key Laboratory of Security and Privacy in Intelligent Transportation, Beijing Jiaotong University, China 
  }
}

\maketitle

\begin{abstract}
Decentralization has been widely acknowledged as a core virtue of blockchains.
However, in the past, there have been few measurement studies on measuring and comparing the actual level of decentralization between existing blockchains using multiple metrics and granularities.
This paper presents a new comparison study of the degree of decentralization in Bitcoin and Ethereum, the two most prominent blockchains, with various decentralization metrics and different granularities within the time dimension.
Specifically, we measure the degree of decentralization in the two blockchains during 2019 by computing the distribution of mining power with three metrics (Gini coefficient, Shannon entropy, and Nakamoto coefficient) as well as three granularities (days, weeks, and months).
Our measurement results with different metrics and granularities reveal the same trend that, compared with each other, the degree of decentralization in Bitcoin is higher, while the degree of decentralization in Ethereum is more stable.
To obtain the cross-interval information missed in the fixed window based measurements, we propose the sliding window based measurement approach.
The corresponding results demonstrate that the use of sliding windows could reveal additional cross-interval information overlooked by the fixed window based measurements, thus enhancing the effectiveness of measuring decentralization in terms of continuous trends and abnormal situations.
We believe that the methodologies and findings in this paper can facilitate future studies of decentralization in blockchains.

\end{abstract}


\section{Introduction}

Recent advances in blockchain technologies are driving the rise of decentralized platforms and applications in various areas such as big data~\cite{Karafiloski2017BlockchainSF,smith2017blockchain}, healthcare~\cite{yue2016healthcare} and Internet of Things~\cite{sagirlar2018hybrid}.
Public blockchains such as Bitcoin\cite{nakamoto2019bitcoin} and Ethereum\cite{buterin2014next} are ledgers of transactions performed by nodes in blockchain networks in a decentralized manner.
For most of the public blockchains, the degree of decentralization of resources that decide who generates blocks is the key metric for evaluating the blockchain decentralization~\cite{gencer2018decentralization,kwon2019impossibility,wu2019information,li2019incentivized}. 
This in turn facilitates further understanding of both security and scalability in a blockchain~\cite{kokoris2018omniledger}.

Intuitively, the case that a few parties possess the majority of the overall resources indicates a more centralized control of blockchain, which is potentially less secure. This is due to the fact that the collusion among these few parties can be powerful enough to perform denial-of-service attacks against targeted blockchain users and even falsify historical data recorded in blockchain.
More concretely, in a Proof-of-Work (PoW) blockchain such as Bitcoin~\cite{nakamoto2019bitcoin}, a miner possessing higher mining power has a better chance of generating the next block. 
In Bitcoin, a transaction is considered to be `confirmed' after six blocks as it is estimated that the probability of creating a longer fork after six blocks to defeat the one containing the `confirmed' transaction is negligible.
However, the assumption is not held when a few miners possess over half of overall mining power in the network, in which case these miners are able to launch the commonly known 51\% attack to control the blockchain and double-spend any amount of cryptocurrency.
Through selfish mining~\cite{eyal2014majority}, the difficulty of performing 51\% attack could be further reduced to the demand of possessing 33\% of overall mining power, indicating even weaker security in less decentralized blockchains.

Some research works in the last few years pointed out that Bitcoin and Ethereum are showing a trend towards centralization~\cite{beikverdi2015trend,tschorsch2016bitcoin,gencer2018decentralization}.
The main reason is that, for the purpose of reducing the variance of income, the majority of miners have joined large mining pools and a small set of mining pools are now actually controlling the blockchains.
Motivated by this observation, recent works have started measuring and evaluating the degree of decentralization in Bitcoin and/or Ethereum in a quantitative manner~\cite{Li2020ComparisonOD, kwon2019impossibility, wu2019information,gencer2018decentralization}.
However, most of the existing measurement studies were collecting blocks generated in a single week and then computing the degree of decentralization based on the distribution of block producers within the selected week.
Hence, their results could only reveal a snapshot state of the degree of decentralization, rather than its continuous shifts.
Besides, some of these studies performed measurements using a single metric as well as a single granularity, making the results lack comprehensiveness.

This paper presents a new comparison study of the degree of decentralization in Bitcoin and Ethereum, the two most prominent blockchains.
Compared with existing works, our research makes the following key contributions:
\begin{itemize}[leftmargin=*]
\item Our measurement study is based on the data of blocks generated throughout the year 2019, thus enabling the quantitative evaluation of the long-term pattern and continuous changes of the degree of decentralization.
\item Our measurements adopt various decentralization metrics and different granularities within the time dimension, thus making the results more comprehensive.
Specifically, we compute the degree of decentralization with three measurement metrics (Gini coefficient~\cite{kwon2019impossibility}, Shannon entropy~\cite{wu2019information}, and Nakamoto coefficient~\cite{srinivasan2017quantifying}) as well as three measurement granularities (days, weeks and months).
Our measurement results with different metrics and granularities reveal the same trend that, compared with each other, the level of decentralization in Bitcoin is higher, while the level of decentralization in Ethereum is more stable.
\item Initially, we measure the degree of decentralization using fixed windows, namely windows of fixed interval duration with no overlaps between two consecutive windows.
However, our further analysis indicates that, when we compute the degree of decentralization per interval duration (e.g. per week) with fixed windows, we may overlook the cross-interval (e.g. cross-week) changes, which may even lead to the unawareness of special or abnormal values of the degree of decentralization.
However, a naive solution of reducing the time interval may make the measurements quite sensitive to occasional elements (e.g., a miner with 10\% of overall mining power mined 10\% of overall blocks within a week but 30\% of overall blocks within a certain day of that week).
To overcome these challenges, we propose the sliding window based measurement approach to capture the cross-interval changes of the degree of decentralization.
Our measurement results demonstrate that the use of sliding windows could reveal additional cross-interval information overlooked by the fixed window based measurements, thus enhancing the effectiveness of measuring decentralization in terms of continuous trends and abnormal situations.
\end{itemize}

The rest of this paper is organized as follows: In Section~2, we present our ways of collecting data, introduce the adopted measurement metrics, and show the preliminary results measured with fixed windows. 
Then, in Section~3, we introduce the sliding window based measurement approach and compare the corresponding measurement results with the ones measured with fixed windows. 
Finally, we discuss the related work in Section~4 and conclude in Section~5.

\section{Data collection and preliminary Measurement}
In this section, we describe our data collection methodology and introduce the adopted measurement metrics.
We then present the preliminary results of the degree of decentralization in Bitcoin and Ethereum measured with fixed windows.

\subsection{Data collection}
We leverage Google BigQuery~\cite{tigani2014google} to collect blockchain data.
Google BigQuery is a web service platform that offers an Interactive Application Programming Interface(API) for developers and researchers to collect and parse various types of data, including eight mainstream Proof-of-Work (PoW) blockchain data.
From block 556,459 to block 610,690, we collected 54,231 Bitcoin blocks produced in 2019.
Similarly, we collected 2,204,650 Ethereum blocks produced in 2019 from block 6,988,615 to block 9,193,265.

\subsection{Measurement metrics}
In this paper, we quantify and compare the degree of decentralization in Bitcoin and Ethereum with three different measurement metrics, namely Gini coefficient~\cite{kwon2019impossibility}, Shannon entropy~\cite{wu2019information}, and Nakamoto coefficient~\cite{srinivasan2017quantifying}.
Next, we will introduce the three measurement metrics respectively.

\subsubsection{Gini coefficient}
The Gini coefficient is often used as a gauge of economic inequality, measuring wealth distribution among a population.
In the scenario of measuring decentralization in blockchains, the Gini coefficient could be used to indicate the inequality of the distribution of mining power among miners~\cite{kwon2019impossibility}. Therefore, we adopt Gini coefficient as the first measurement metric, which could be computed by:
\begin{equation}
G=\frac{\sum_{ A_{i},A_{j}\in{A}}|NB_{A_{i}}-NB_{A_{j}}|}{2|A|\sum_{NB_{A_{j}}\in{NB}}{NB_{A_{j}}}}
\label{Gini}
\end{equation}
where $NB_{A_{i}}$ refers to the number of blocks generated by each block producer $A_{i}$ and $NB=\{NB_{A_{i}}|A_{i}\in{A}\}$ and $A$ denote the corresponding sets.
If the deviation of $NB$ is small, the Gini value will be close to 0. Otherwise, the value will be close to 1.
Intuitively, a lower Gini coefficient means that more miners need to collude to compromise a blockchain system, thus indicating a higher degree of decentralization.



\subsubsection{Shannon Entropy}
In 1948, Shannon pointed out that information has redundancy, and the amount of redundancy depends on the distribution of probabilities, or uncertainty, within the information, which could be quantified by the notion of Shannon entropy~\cite{wu2019information}.
We adopt the Shannon entropy as the second measurement metric to measure the degree of randomness and disorder of the distribution of the amount of blocked mined by miners via:
\begin{equation}
    p_i=\frac{b_i}{-\sum_{i=1}^{n}b_i}
\end{equation}
\begin{equation}
    E=-\sum_{i=1}^{n}p_{i}log_{2}p_{i}
\end{equation}
where $b_i$ denotes the number of blocks generated by miners and n denotes the range of miners that entropy is computed for. 
Intuitively, a higher Shannon entropy means a higher degree of randomness of the distribution of mining power, thus indicating a higher degree of decentralization.
\subsubsection{Nakamoto coefficient}
The aforementioned two metrics are a bit abstract and the degree of decentralization quantified via them is not directly associated with the security in blockchains.
Therefore, we adopt the Nakamoto coefficient as the third measurement metric in this paper.
Nakamoto coefficient is defined as the minimum number of entities required to collude for gathering over 51\% of the overall mining power to compromise a blockchain system~\cite{srinivasan2017quantifying}, which could be computed via:
\begin{equation}
    N=min\{k\in[1,\cdots,K]:\sum_{i=1}^{k}p_{i}\geq0.51\}
\end{equation}
Intuitively, a higher Nakamoto coefficient means more miners (or mining pools) need to combine their mining power to reach the 51\% threshold to take over the blockchain.
Compared with the Gini coefficient, the Nakamoto coefficient 
is more intuitive for revealing the association between the degree of decentralization and security in blockchains.

\subsection{Preliminary Measurement}
Next, we present the preliminary results of the degree of decentralization in Bitcoin and Ethereum measured using fixed windows with lengths of a day, a week and a month, respectively.
Here, fixed windows refer to windows of fixed interval duration. There will be no overlaps between two consecutive fixed widows.
In other words, we show the daily, weekly and monthly degree of decentralization in the two blockchains.

\subsubsection{Measurement results in Bitcoin}
\begin{figure}[t!]
    \centering
 \includegraphics[width=1\columnwidth]{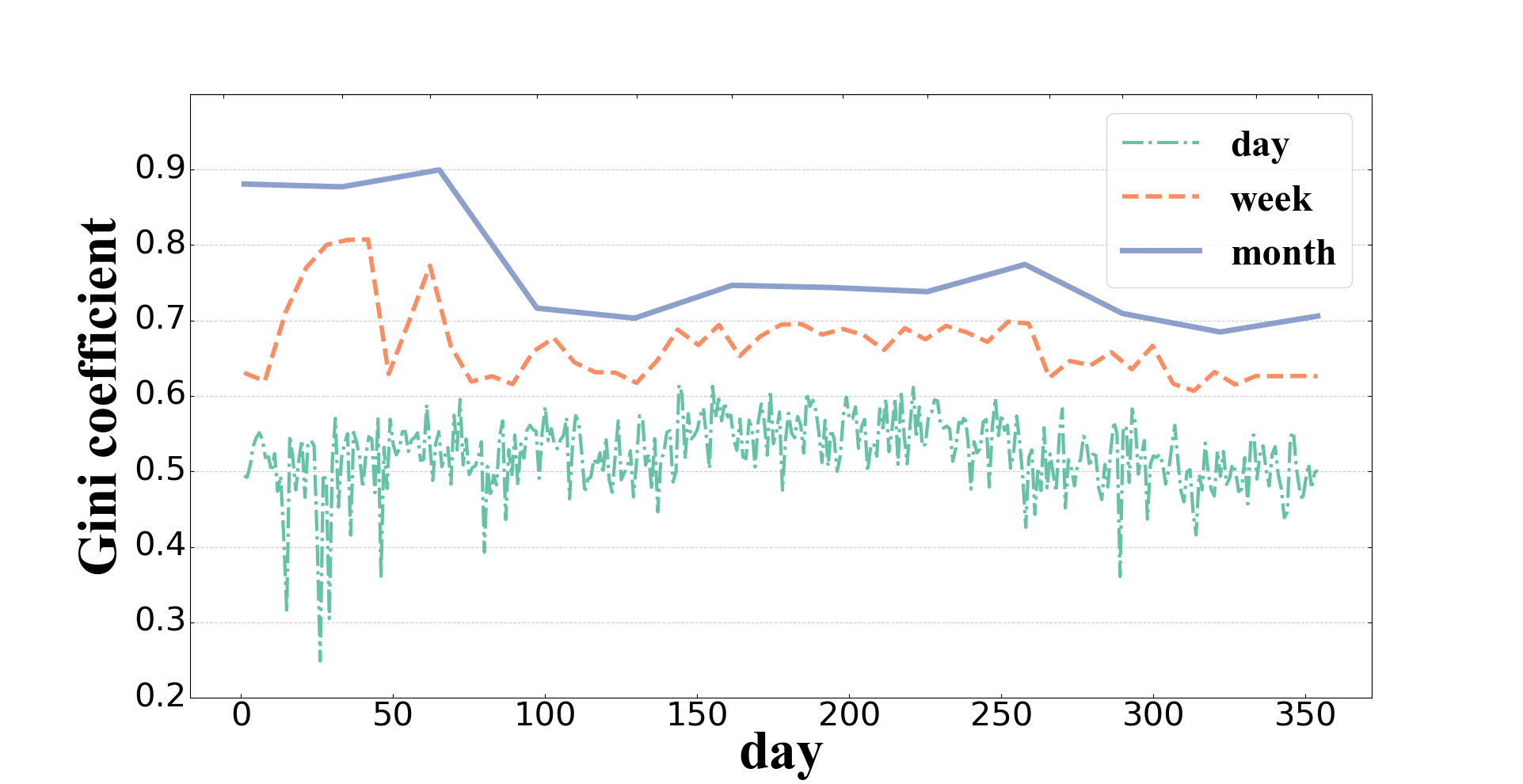}
    \caption{Gini coefficient measured in Bitcoin using fixed windows}
    \label{BtcFixGini} 
\end{figure}

\paragraph{Gini coefficient}
Fig.~\ref{BtcFixGini} presents the Gini coefficient measured in Bitcoin using fixed windows.
As can be seen from the results, the monthly Gini coefficients, with the highest values close to 0.90 during the first three months, are always higher than the daily and weekly ones.
The variation tendency of the weekly Gini coefficients is similar to that of the monthly Gini coefficients.
Finally, the daily Gini coefficients are much lower than the monthly and weekly ones in general.
Most of the daily Gini coefficients are within the range of 0.45 to 0.60, while some extreme values could reach around 0.25 during the first three months.


\begin{figure}[htbp]
    \centering
    \includegraphics[width=1\columnwidth]{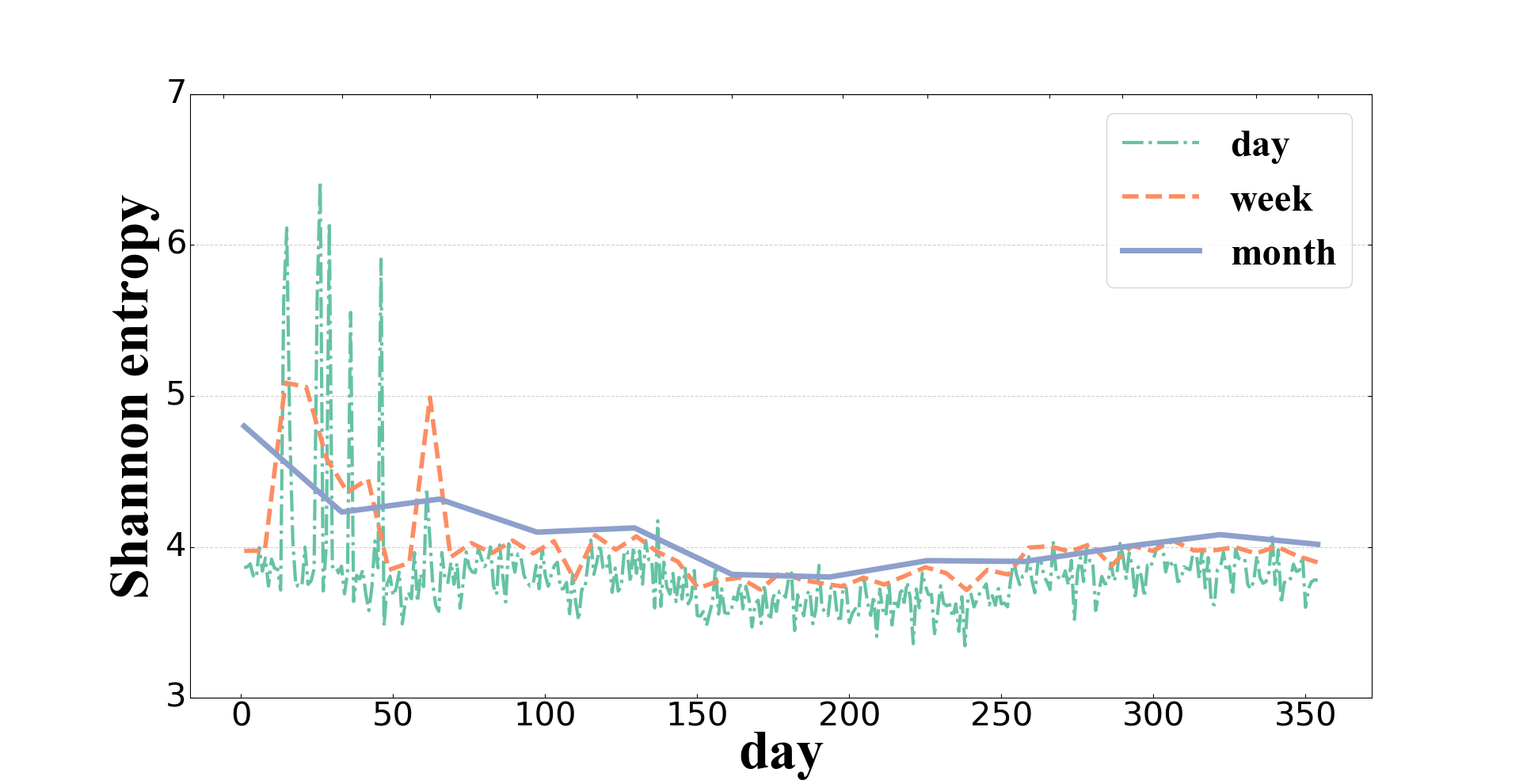}
    \caption{Shannon entropy measured in Bitcoin using fixed windows}
    \label{BtcFixEntropy}
\end{figure}

\paragraph{Shannon entropy}
Next, Fig.~\ref{BtcFixEntropy} presents the Shannon entropy measured in Bitcoin using fixed windows.
From the results, we can see that the overall patterns of the daily, weekly and monthly Shannon entropy are quite close.
We can also observe that during the first two months, the value of Shannon entropy is higher, regardless of the selected measurement granularity.
Specifically, the daily Shannon entropy is within the range of 3.5 to 4.0, while some extreme values could become higher than 5.5.


\begin{figure}[htbp]
    \centering
    \includegraphics[width=1\columnwidth]{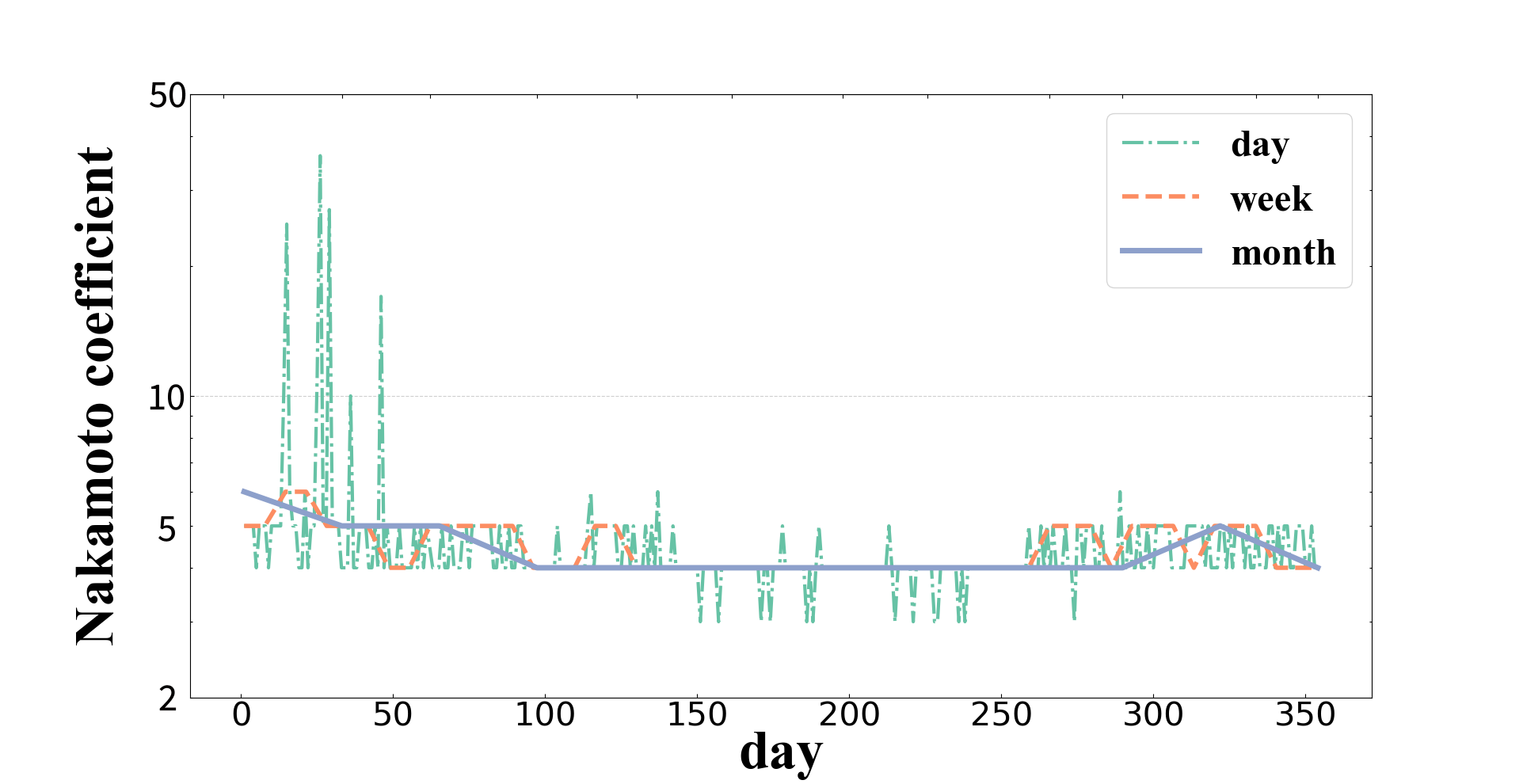}
    \caption{Nakamoto coefficient measured in Bitcoin using fixed windows}
    \label{BtcFixA}
\end{figure}

\paragraph{Nakamoto coefficient}
Finally, Fig.~\ref{BtcFixA} presents the Nakamoto coefficient measured in Bitcoin using fixed windows.
The results show that from day 100 to day 260, the Nakamoto coefficients measured with all the three granularities are relatively stable at 4. 
Outside this range, the Nakamoto coefficient mainly oscillates between 4 and 5, while the highest values of daily Nakamoto coefficient appeared during the first 50 days could be higher than 35.


\paragraph{Summary}
By observing the three figures together, we could find that the degree of decentralization measured with all the three metrics tends to be higher and with more fluctuations during the first 50 days while to be lower and more stable during the rest of the year.
To figure out the underlying reason, we investigated the detailed block data.
We take day 14, namely Jan. 14th in 2019, as an example.
On day 14, the daily Gini coefficient is 0.34 and the daily Shannon entropy is 6.2, both of which are extreme values.
By investigating the block data, we found two abnormal blocks.
Specifically, the no. 558,473 and no. 558,545 blocks contain more than 80 independent coinbase addresses and more than 90 independent coinbase addresses, respectively, resulting in each of these two blocks being associated with more than 80 blocks producers.
Because of this, day 14 has only 148 blocks created on that day but is with an extremely large set of miners, resulting in a very small daily Gini coefficient as well as a very large daily Shannon entropy.


\subsubsection{Measurement results in Ethereum}
\begin{figure}[t!]
    \centering
    \includegraphics[width=1\columnwidth]{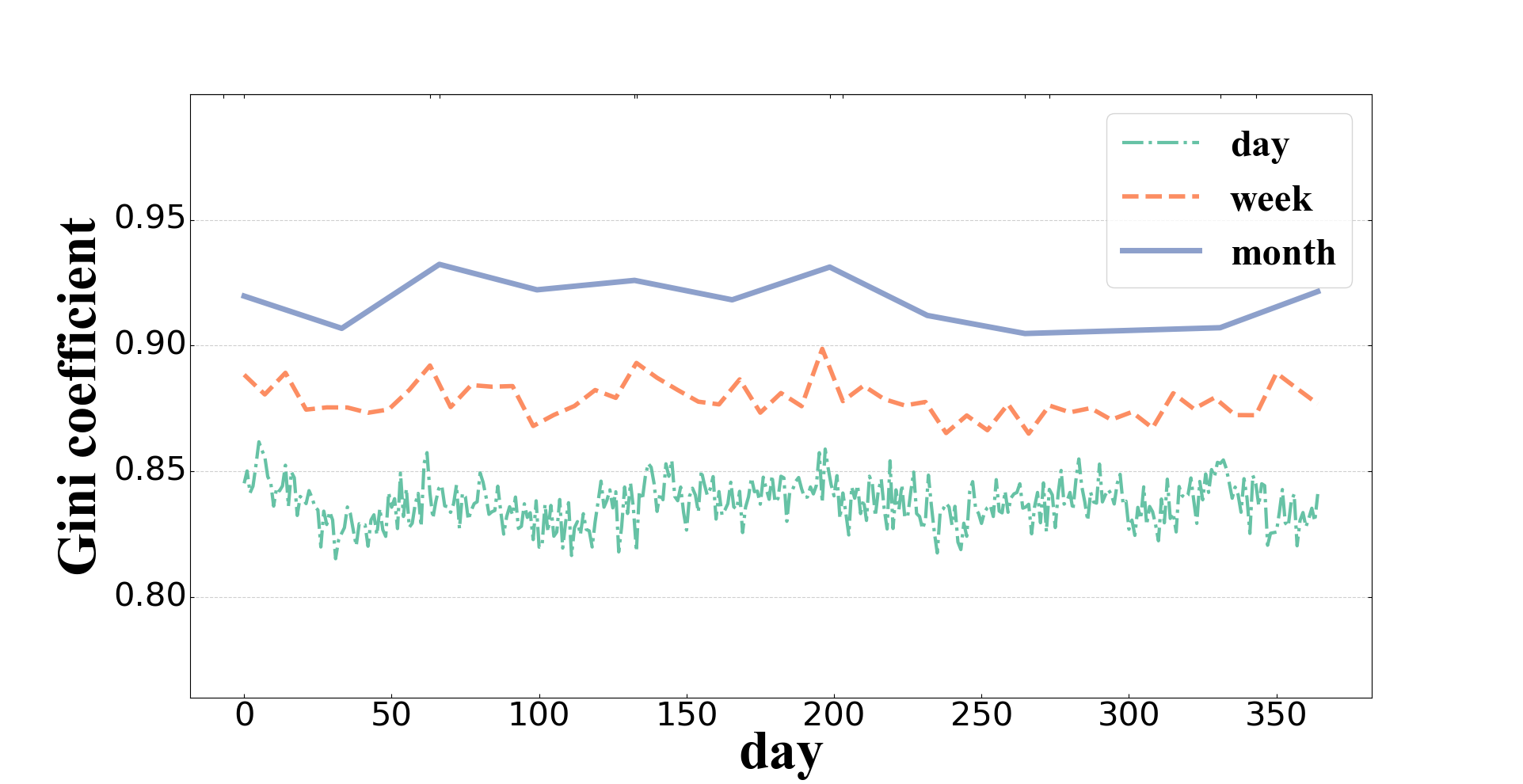}
    \caption{Gini coefficient measured in Ethereum using fixed windows}
    \label{EthFixGini}
\end{figure}

\paragraph{Gini coefficient}
Fig.~\ref{EthFixGini} presents the Gini coefficient measured in Ethereum using fixed windows.
We could observe the same trends as revealed in the measurement results in Bitcoin.
That is, the monthly Gini coefficients are always higher than the daily and weekly ones and the daily Gini coefficients are much lower than the monthly and weekly ones in general.
Besides, compared with those in Bitcoin, the Gini coefficients in Ethereum are higher and also more stable.

\begin{figure}[t!]
    \centering
    \includegraphics[width=1\columnwidth]{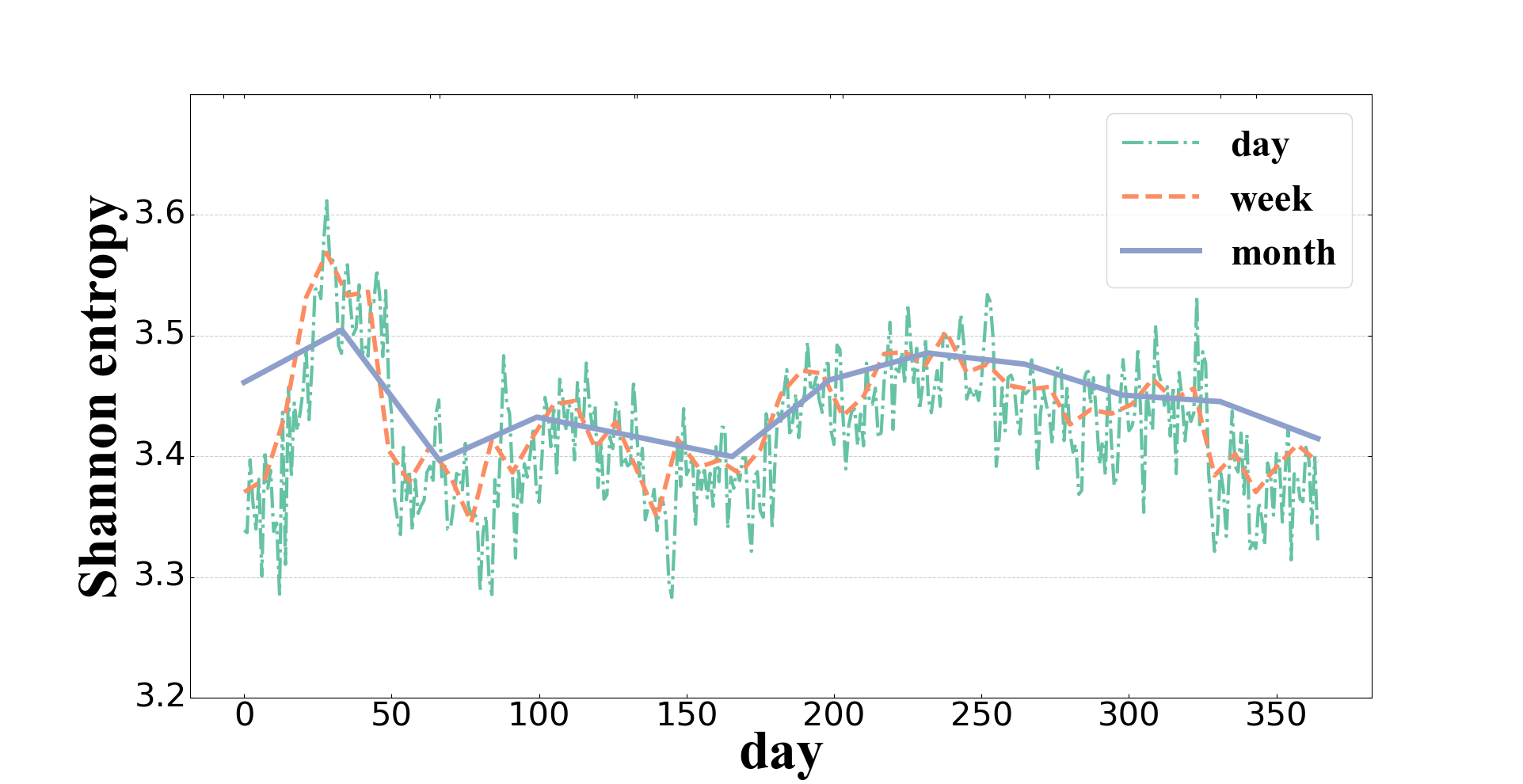}
    \caption{Shannon entropy measured in Ethereum using fixed windows}
    \label{EthFixEntropy}
\end{figure}

\paragraph{Shannon enrtopy}
As shown in Fig.~\ref{EthFixEntropy}, the trends of the Shannon entropy measured with different granularities are roughly the same.
More concretely, most of the values of Shannon entropy are within the range of 3.3 to 3.5.

\begin{figure}[t!]
    \centering
    \includegraphics[width=1\columnwidth]{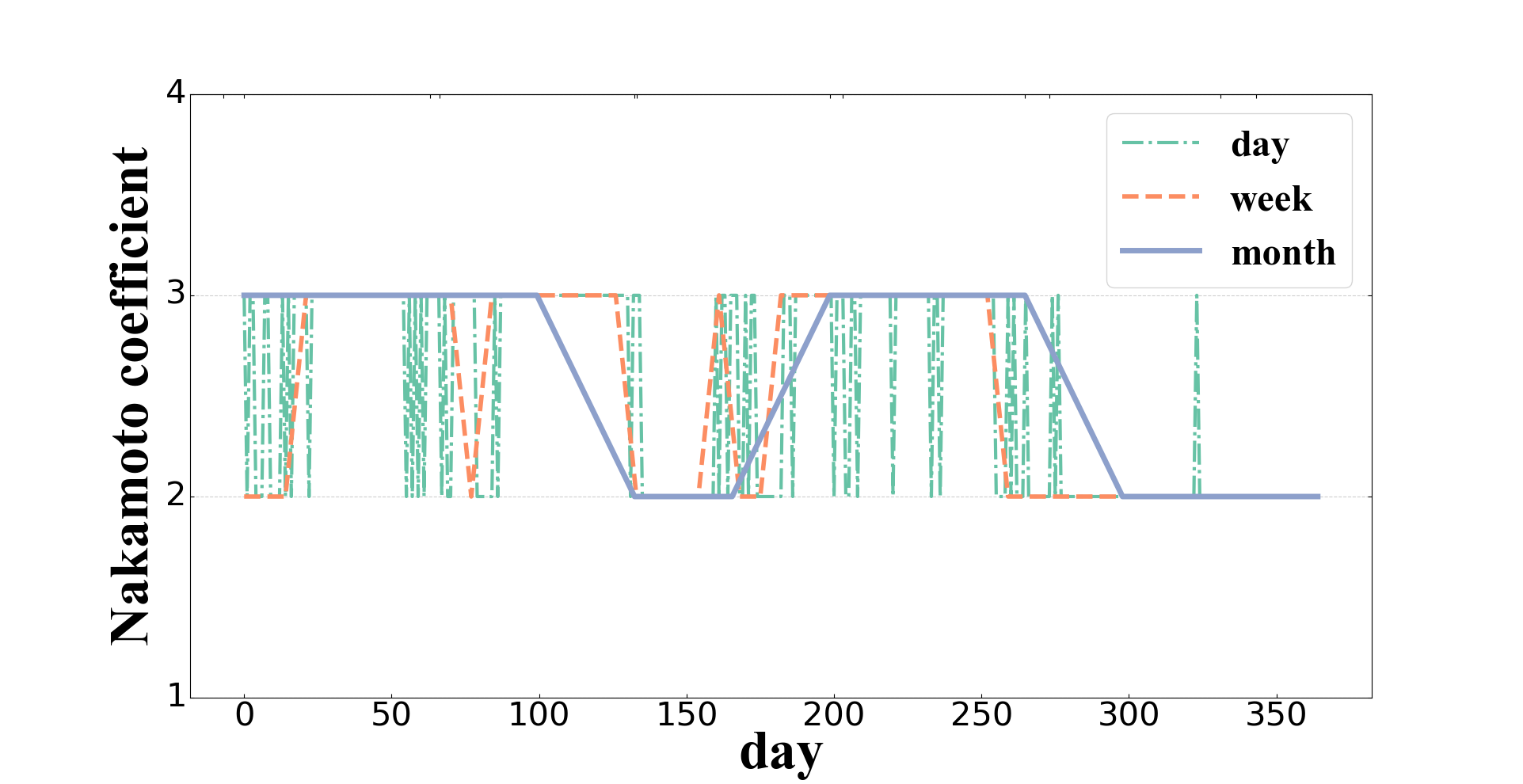}
    \caption{Nakamoto coefficient measured in Ethereum using fixed windows}
    \label{EthFixA}
\end{figure}

\paragraph{Nakamoto coefficient}
Next, Fig.~\ref{EthFixA} presents the Nakamoto coefficient measured in Ethereum using fixed windows.
We could see that the Nakamoto coefficients measured with different granularities are quite stable, which just fluctuates between 2 and 3.


\paragraph{Summary}
Overall, the degree of decentralization in Ethereum measured with all the three metrics is very stable. There is no abnormal value observed during the year.

\subsubsection{Bitcoin versus Ethereum}
To sum up, we found that during the year 2019, our measurement results with different metrics and granularities reveal the same trend that, compared with each other, the level of decentralization in Bitcoin is higher, while the level of decentralization in Ethereum is more stable.

\begin{figure}[t!]
     \centering
     \includegraphics[width=3in]{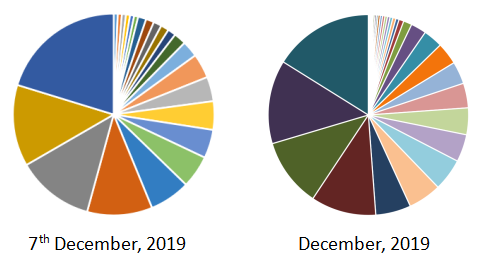}
     \caption{The distribution of blocks produced in Bitcoin within a day and a month}
     \label{DayDistribution}
 \end{figure}

Besides, after comprehensively measuring the degree of difference in Bitcoin and Ethereum with fixed windows, we could see that the change of measurement granularities does not significantly impact the overall trends of Shannon entropy and Nakamoto coefficient.
However, the ranges of Gini coefficients measured with different granularities are quite different.
We would like to explain the underlying reason for the phenomenon based on the formula of the Gini coefficient, namely Eq.~\ref{Gini}.
We could take the day 7th December 2019 and the corresponding month December 2019 as an example.
As shown in Fig.~\ref{BtcFixEntropy}, the daily Shannon entropy measured on the aforementioned day and the monthly Shannon entropy measured in the aforementioned month are quite close.
As revealed by Fig.\ref{DayDistribution}, the changes of the ratio of blocks mined by top miners in the two pie charts are not substantial.
However, a longer length of fixed windows could take huge numbers of miners who may create only a few blocks into the computation.
As a result, the population of top miners remains the same while the population of bottom miners becomes larger, resulting in a higher Gini coefficient.



So far, our measurements are based on fixed windows, namely windows of fixed interval duration.
However, our further analysis indicates that, when we compute the degree of decentralization per interval duration (e.g. per week) with fixed windows, we may overlook cross-interval (e.g. cross-week) changes, which may even lead to the unawareness of special or abnormal values of the degree of decentralization.
To resolve the problem, in the next section, we propose the sliding window based measurement approach to capture the cross-interval changes of the degree of decentralization.


 
\section{Measurement with sliding windows}
In this section, we propose the sliding window based measurement approach to replace the fixed windows used in the last section so that the results could be finer-grained and the cross-interval changes could be captured. 
We first present our measuring methodology of using sliding windows and then the corresponding results.

\begin{figure}[t!]
     \centering
     \includegraphics[width=1\columnwidth]{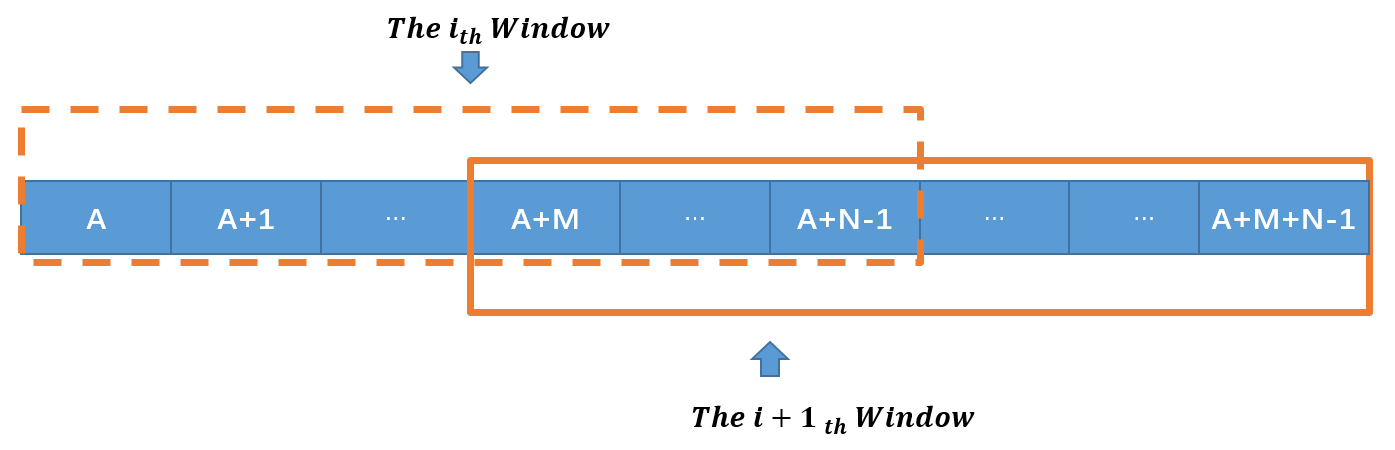}
     \caption{Sliding window}
     \label{Sliding window}
 \end{figure}

\subsection{Sliding Window}\label{AA}
As we have described in Section~2, the use of fixed windows may make the results overlook the cross-interval changes and hence, miss important signals such as special or abnormal values of the degree of decentralization.
For instance, a miner may dominantly produced most of blocks during four consecutive days, which should have been considered as an unusual sign. However, if the four consecutive days are the last two days of the first week and the first two days of the second week, the values of the two corresponding weekly degree of decentralization may not be high enough to attract attention.
A naive solution for this problem is to reduce the length of the fixed windows to capture finer-grained changes.
However, this naive solution may make the measurements quite sensitive to occasional elements.
As an example, it is possible that a miner with 10\% of overall mining power mined 10\% of overall blocks within a week but 30\% of overall blocks within a certain day of that week.
Then, even though the amount of blocked mined by this miner during the week exactly matches the mining power owned by the miner, the degree of decentralization measured with short fixed windows may fluctuate sharply.
To overcome these challenges, we propose the sliding window based measurement approach to capture the cross-interval changes of the degree of decentralization without having to reduce the window size.

We now present the setting of sliding windows with more details.
First, based on the 10 minutes per block production rate in Bitcoin, we could approximate the numbers of blocks produced in Bitcoin per day, per week and per month, respectively.
Then, based on the results, we could set the corresponding window sizes used in Bitcoin to be 144, 1008, and 4320, respectively. 
Similarly, based on the fact that Ethereum generates 6,000 blocks per day on average, we could set the corresponding window sizes used in Ethereum to be 6,000, 42,000 and 180,000, respectively.
After that, by denoting the length of a sliding window as $N$ and the sliding step length as $M$, we can obtain $L$ measurement results when the total number of measured blocks is $S$, as defined below:
\begin{equation}
    L = \frac{S-N}{M} + 1  \label{Length}
\end{equation}
In this paper, in order to simplify and clarify the main point of the sliding window based measurement approach, we take the sliding step length $M$ half of the size of the sliding window $N$, so the number of times of computing the degree of decentralization in 2019 could be doubled.



As shown in Fig.~\ref{Sliding window}, when the window size is $N$, assuming that the first block within the $i_{th}$ window has block number $A$, the block number of the last block within the $i_{th}$ window would be $A+N-1$. 
Then, given the sliding step length $M$, the first and last blocks within the  ${i+1}_{th}$ window would have the block numbers \textit{A+M} and \textit{A+M+N-1}, respectively.
As a result, we can see that the two consecutive windows share an amount of $N-M$ overlapping blocks, so the cross-interval changes could be captured.


\begin{figure}[t!]
     \centering
     \includegraphics[width=1\columnwidth]{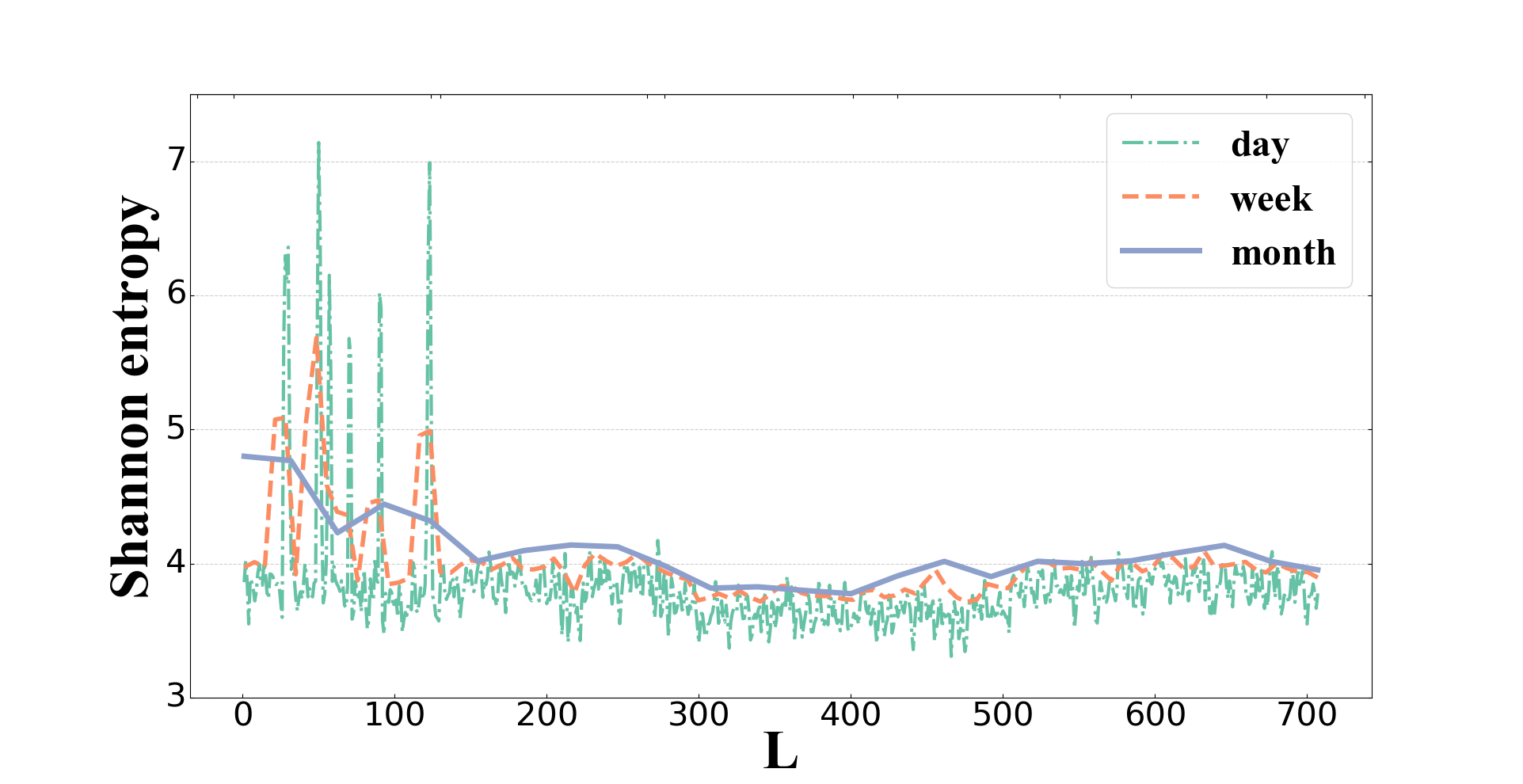}
     \caption{Shannon entropy measured in Bitcoin using sliding windows}
     \label{BtcStepEntropy}
 \end{figure}
 
  \begin{figure}[t!]
     \centering
     \includegraphics[width=1\columnwidth]{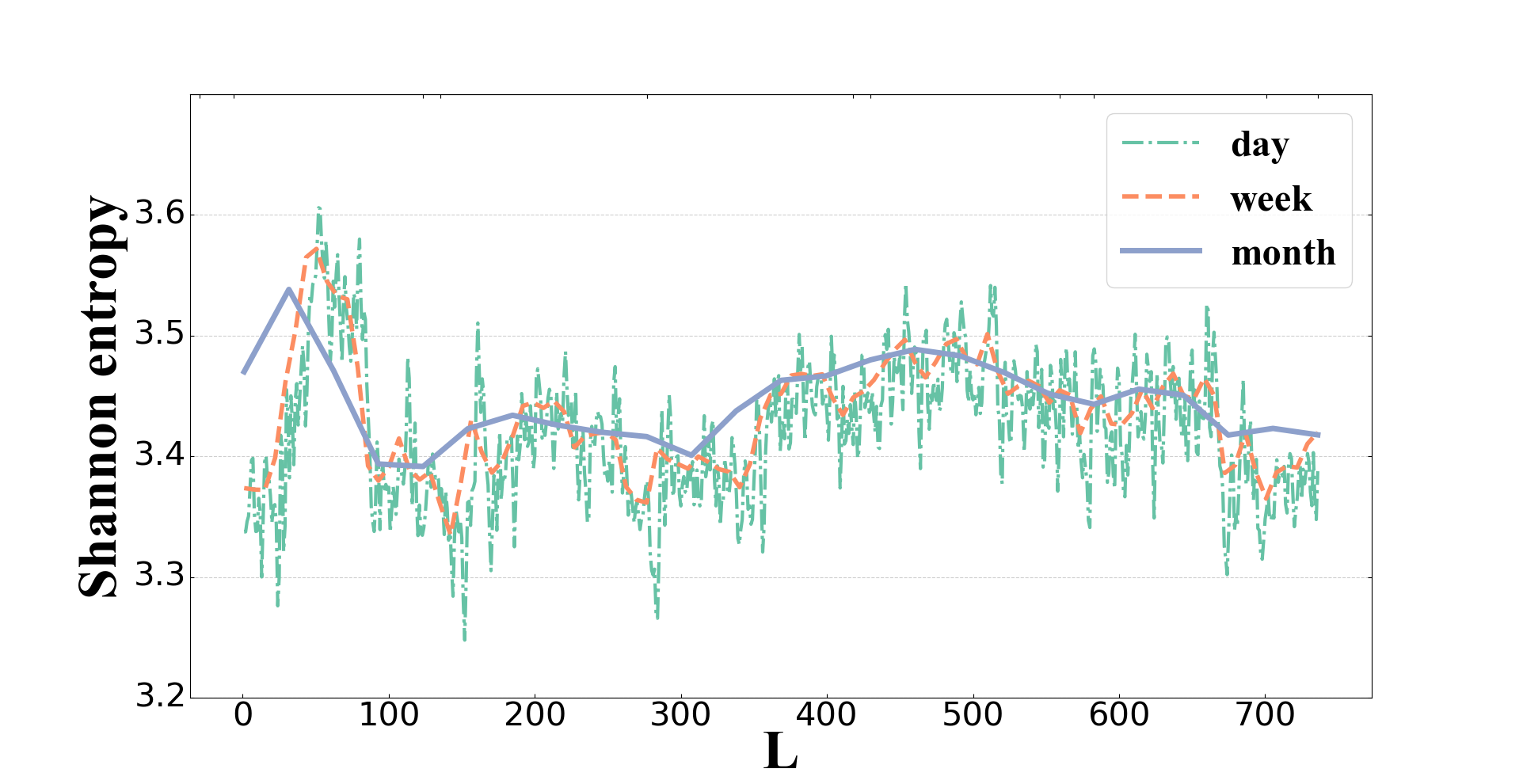}
     \caption{Shannon entropy measured in Ethereum using sliding windows}
     \label{EthStepEntropy}
 \end{figure}

\subsection{Measurement Result}\label{BB}
Based on the aforementioned sliding window based measurement approach, Fig.~\ref{BtcStepEntropy} shows the Shannon entropy measured in Bitcoin with the window size changed from 144, the number of blocks generated per day, to 1008, blocks produced per month and finally to 4230, blocks created per year.
According to Eq.~\ref{Length}, with the window size close to the number of blocks generated per day, we can get about 700 results using sliding windows instead of 365 results using fixed windows.
That is, the number of results is almost doubled.
From the perspective of statistical results, we can see that the average values of Shannon entropy obtained by sliding windows and fixed windows are quite close.  
Specifically, the average values of Shannon entropy measured with one-day, one-week and one-month long sliding windows are about 3.810, 4.002 and 4.091, respectively.


Compared with Fig.~\ref{BtcFixEntropy} that shows the Shannon entropy measured with fixed windows, most of the values of Shannon entropy measured with one-day long sliding windows locate in the range of 3.5 to 4.0.
Also, we can see that the sliding window based results reveal more extreme values (i.e., $> 5.0$) than the ones revealed from the fixed window based results.
Moreover, we notice that the abnormal changes of Shannon entropy revealed in Fig.~\ref{BtcFixEntropy} have been magnified in Fig.~\ref{BtcStepEntropy}.


As for the results measured with one-week long sliding windows, we can observe that during the range of $N$ from 40 to 100, which approximately equals to the range of day from 20 to 50, the one-week long fixed window based results only show a continuous downward trend while the one-week long sliding window based results can reveal more cross-interval changes.
Therefore, by replacing fixed windows with sliding windows, it would be easier for us to discover special or abnormal changes of the degree of centralization in a more timely manner.




Next, Fig.~\ref{EthStepEntropy} presents the Shannon entropy measured in Ethereum using sliding windows.
By comparing the results in Fig.~\ref{EthStepEntropy} and Fig.~\ref{EthFixEntropy}, we can find that the sliding window based results are quite close to the fixed window based ones.
More concretely, the degree of decentralization in Ethereum measured with sliding windows also shows a stable trend, and most of the measured results are between 3.3 and 3.5. 
From the perspective of statistical results, the average values of Shannon entropy measured with one-day, one-week and one-month long sliding windows are about 3.420, 3.433 and 3.445, respectively.
Again, we can see that, compared with Bitcoin, Ethereum tends to be more stable but less decentralized in terms of the measurements of the Shannon entropy.

\begin{figure}[t!]
     \centering
     \includegraphics[width=3in]{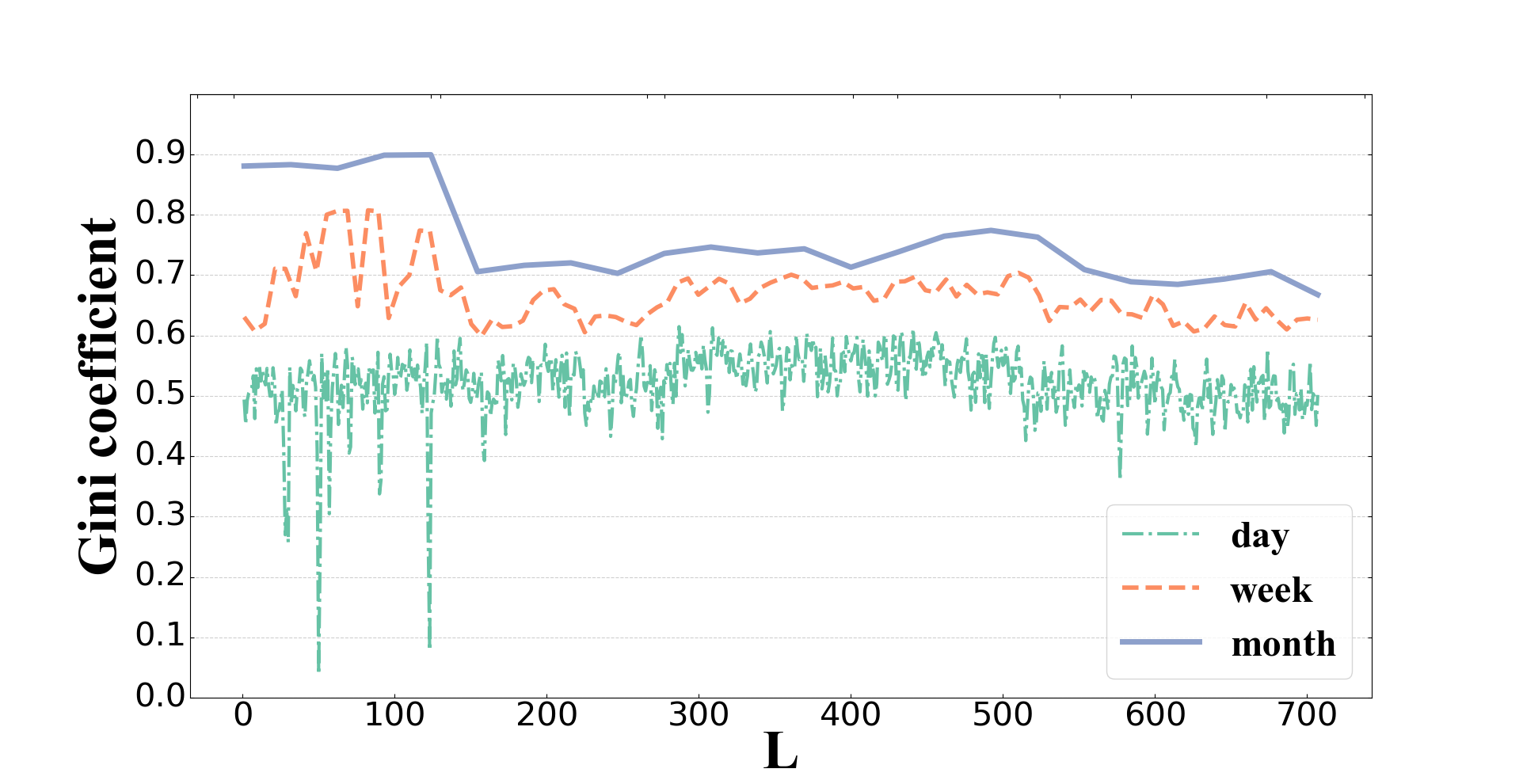}
     \caption{Gini coefficient measured in Bitcoin using sliding windows}
     \label{BtcStepGini}
 \end{figure}
\begin{figure}[t!]
     \centering
     \includegraphics[width=3in]{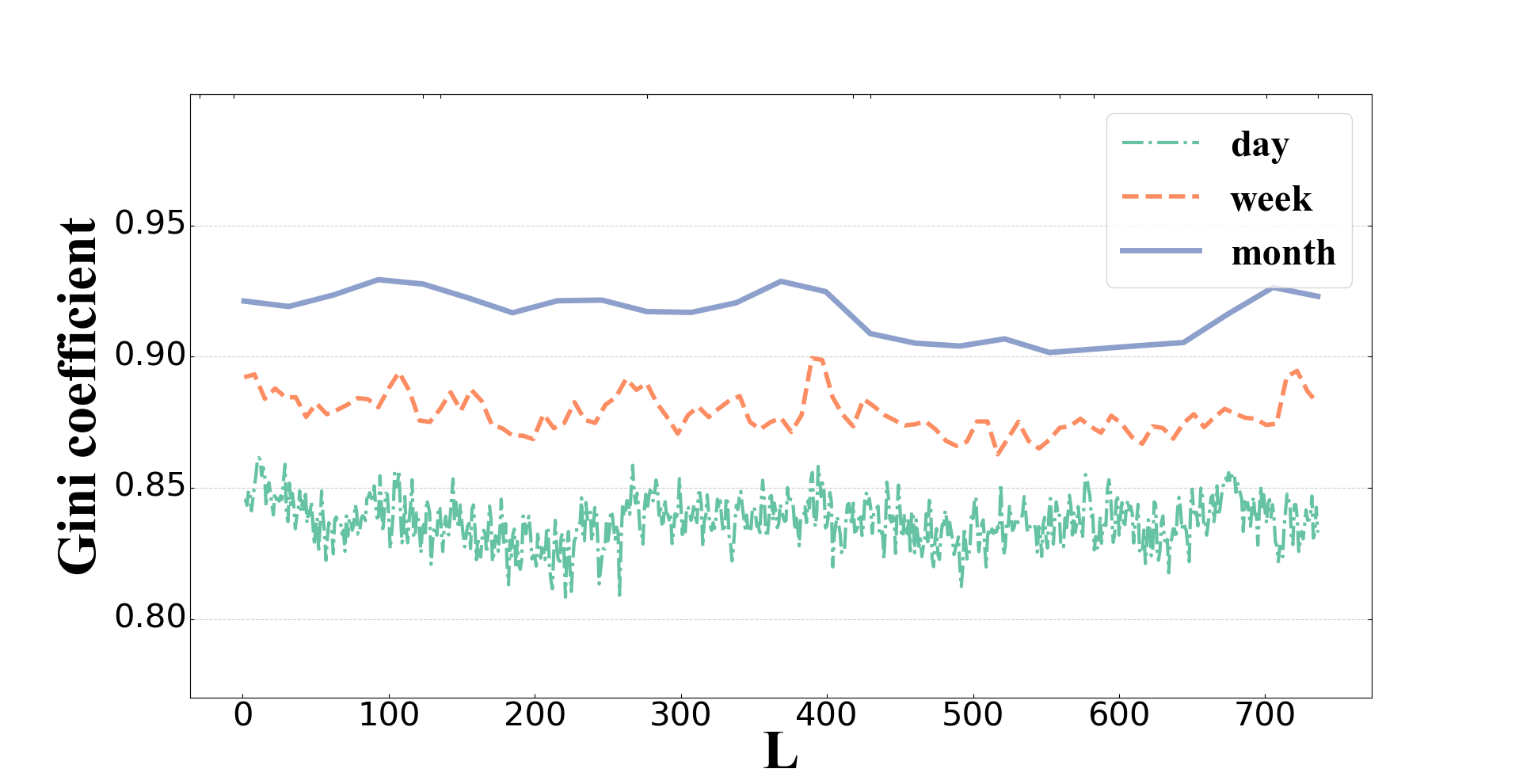}
     \caption{Gini coefficient measured in Ethereum using sliding windows}
     \label{EthStepGini}
 \end{figure}

Following the same methodology, we now observe the Gini coefficient measured in Bitcoin and Ethereum using sliding windows.
The results are shown in Fig.~\ref{BtcStepGini} and Fig.~\ref{EthStepGini}, respectively.
For Bitcoin (Fig.~\ref{BtcStepGini}), similar to the trends revealed in Fig.~\ref{BtcFixGini}, the measured values of the Gini coefficient are highly correlated with the measurement granularities, meaning that larger granularities could result in higher measured values.
Specifically, the average values of the Gini coefficient measured with one-day, one-week and one-month long sliding windows are about 0.523, 0.667 and 0.760, respectively.
Also, similar to what we have observed from the measurements of Shannon entropy, the use of sliding windows could reveal additional cross-interval information overlooked in the fixed window based measurements, which could better reflect the abnormality of the changes of degree of decentralization.  
For Ethereum (Fig.~\ref{EthStepGini}), similar to the trends revealed in Fig.~\ref{EthFixGini}, the measured values of the Gini coefficient are quite stable.
The average values of the Gini coefficient measured with one-day, one-week and one-month long sliding windows are about 0.837, 0.878 and 0.916, respectively.
Compared with Bitcoin, Ethereum tends to be significantly less decentralized in terms of the measurements of the Gini coefficient, which might be affected by the huge difference of the block production rates between Bitcoin and Ethereum.


\begin{figure}[t!]
     \centering
     \includegraphics[width=3in]{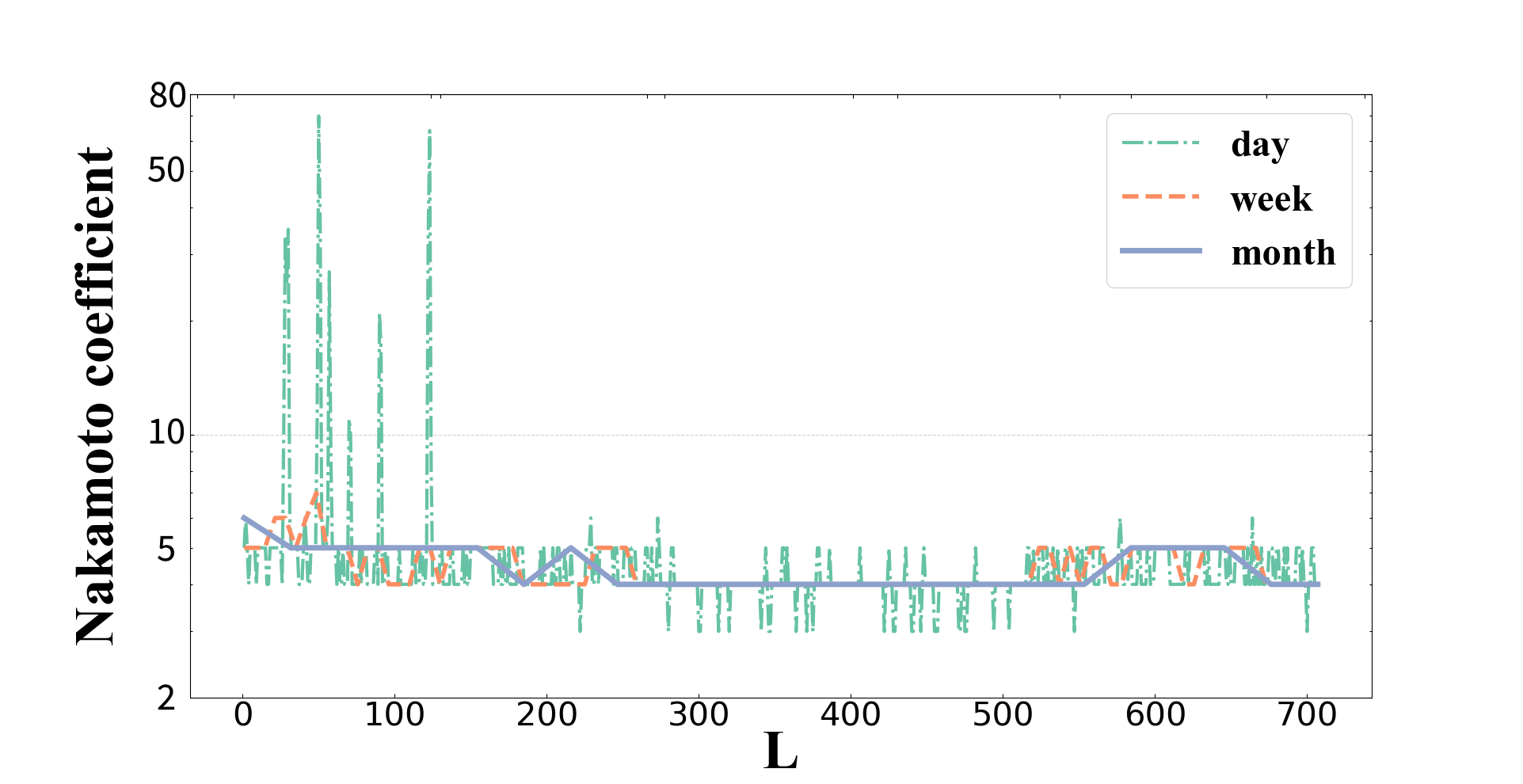}
     \caption{Nakamoto coefficient measured in Bitcoin using sliding windows}
     \label{BtcASteped}
 \end{figure}

\begin{figure}[t!]
     \centering
     \includegraphics[width=3in]{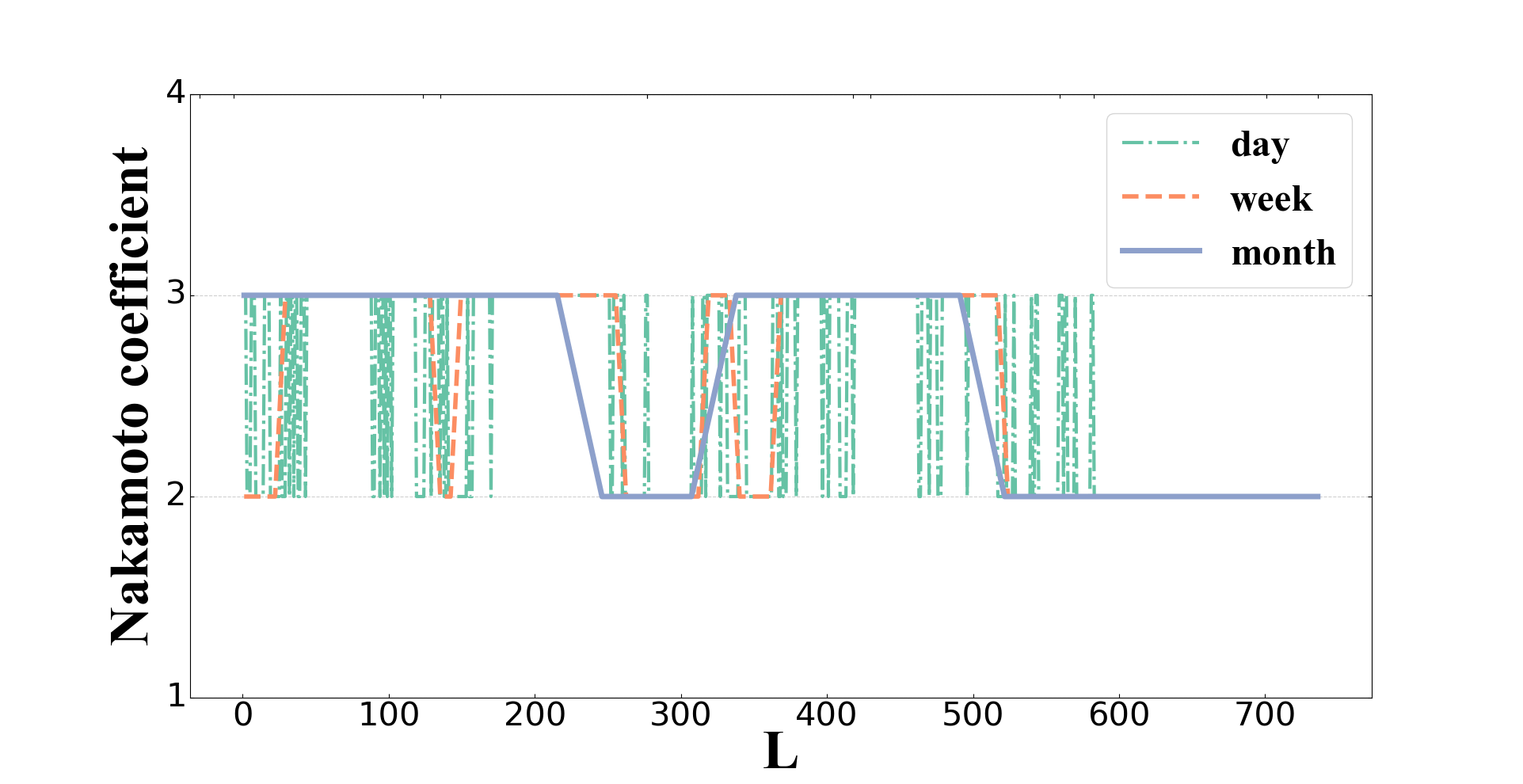}
     \caption{Nakamoto coefficient measured in Ethereum using sliding windows}
     \label{EthAStep}
 \end{figure}

Finally, we observe the Nakamoto coefficient measured in Bitcoin and Ethereum using sliding windows.
The results are shown in Fig.~\ref{BtcASteped} and Fig.~\ref{EthAStep}, respectively.
For Bitcoin (Fig.~\ref{BtcASteped}), compared with the results in Fig.~\ref{BtcFixA}, most of the measured values of the Nakamoto coefficient are between 4 and 5. 
We can also see that some extreme values measured with fixed windows have been doubled in the results measured with one-day long sliding windows.
Moreover, at the moment that $N$ equals to 120 (i.e., day 60), we can find that the abnormal change of the Nakamoto coefficient can be clearly observed in the sliding window based results, but not in the fixed window based results, which proves that sliding window based approach is able to discover changes missed by the fixed window based approach.
For Ethereum (Fig.~\ref{EthAStep}), we can see that the majority of the measured values of the Nakamoto coefficient are between 2 and 3, indicating that the majority of mining power in Ethereum is dominantly controlled by a few entities. 
Similar to the conclusion revealed by other metrics, compared with Bitcoin, Ethereum tends to be less decentralized, as revealed by the results of the Nakamoto coefficient.

To sum up, by comparing the sliding window based results with the fixed window based results using different metrics and granularities, we can conclude that the sliding window based measurement approach can reveal additional cross-interval information overlooked by the fixed window based measurements, thus facilitating the observation of special and abnormal changes of the degree of decentralization in blockchains.

\section{RELATED WORK}
Most of the recent works on decentralization in blockchain have focused on Bitcoin~\cite{6824541,wang2019measurement,Miller2015DiscoveringB,eyal2014majority,eyal2015miner}. 
These works pointed out that Bitcoin shows a trend towards centralization because of the emergence of mining pools.
In~\cite{eyal2014majority}, authors proposed the selfish mining, which reduces the bar of performing 51\% attack to possessing over 33\% of computational power in Bitcoin.
Later, authors in~\cite{eyal2015miner} analyzed the mining competitions among mining pools in Bitcoin from the perspective of game theory and proposed that a rational mining pool may get incentivized to launch a block withholding attack to another mining pool.
From the perspective of mining pools, authors in~\cite{wang2019measurement} tracked more than 1.56 million blocks (including about 257 million historical transactions) and found that a few mining pools was controlling and will keep controlling most of the computing resources of the Bitcoin network. 
Through the research of a small fraction of the network, authors in~\cite{Miller2015DiscoveringB} found that the nodes connected with the major mining pools occupy higher mining capacity. 
Authors in~\cite{6824541} analyzed the centralization of Bitcoin in Bitcoin Web Wallets, Protocol Maintenance, BlockChain Forks, and proposed some possible ways to enhance the decentralization.

Besides, recently, there have been a few studies on comparing the degree of decentralization between Bitcoin and Ethereum~\cite{gencer2018decentralization,wu2019information}.
Relying on a new measurement technology that could be used to obtain the application layer information, the work in~\cite{gencer2018decentralization} focused on the Falcon network and mature Internet measurement technology application and found that (1) Bitcoin has a higher capacity network than Ethereum, but Bitcoin may have more clustered nodes in datacenters and 
(2) Bitcoin and Ethereum have fairly centralized mining processes that require future research to further decentralize the consensus protocol. 
The work in~\cite{wu2019information} compared the degree of decentralization of Bitcoin and Ethereum with Shannon entropy and concluded that Bitcoin is usually more decentralized than Ethereum.
To the best of our knowledge, our paper is the first research work that quantifies the degree of decentralization in Bitcoin and Ethereum with multiple metrics and granularities using both fixed and sliding windows.


\section{Conclusion}
In this paper, we present a new comparison study and analysis of the degree of decentralization in the two most prominent blockchain, Bitcoin and Ethereum, the two most prominent blockchains, with various decentralization metrics and different granularities within the time dimension.
We quantified the degree of decentralization in the two blockchains during 2019 by computing the distribution of mining power with three metrics (Gini coefficient, Shannon entropy, and Nakamoto coefficient) as well as three granularities (days, weeks, and months).
Our measurement results reveal that, compared with each other, the degree of decentralization in Bitcoin is higher, while the degree of decentralization in Ethereum is more stable.
In our further analysis, we propose the sliding window based measurement approach to capture the cross-interval changes s of the degree of decentralization missed in the fixed window based measurements. 
Our measurement results demonstrate that the use of sliding windows could reveal additional cross-interval information overlooked by the fixed window based measurements, thus enhancing the effectiveness of measuring decentralization in terms of continuous trends and abnormal situations. 
We believe that the methodologies and findings in this paper can facilitate future studies of decentralization in blockchains.	

\section*{Acknowledgement}
The work in this paper is supported by Fundamental Research Funds for the Central Universities (No. 2019RC038).
Chao Li (Email: li.chao@bjtu.edu.cn) is the corresponding author of this paper.

\renewcommand\refname{Reference}

\bibliographystyle{plain}

\bibliography{main.bib}

\begin{thebibliography}{10}

\bibitem{beikverdi2015trend}
Alireza Beikverdi and JooSeok Song.
\newblock Trend of centralization in bitcoin's distributed network.
\newblock In {\em 2015 IEEE/ACIS 16th SNPD}, pages 1--6. IEEE, 2015.

\bibitem{buterin2014next}
Vitalik Buterin et~al.
\newblock A next-generation smart contract and decentralized application
  platform.
\newblock {\em white paper}, 3:37, 2014.

\bibitem{eyal2015miner}
Ittay Eyal.
\newblock The miner's dilemma.
\newblock In {\em 2015 IEEE Symposium on Security and Privacy}, pages 89--103.
  IEEE, 2015.

\bibitem{eyal2014majority}
Ittay Eyal and Emin~G{\"u}n Sirer.
\newblock Majority is not enough: Bitcoin mining is vulnerable.
\newblock In {\em International conference on financial cryptography and data
  security}, pages 436--454. Springer, 2014.

\bibitem{gencer2018decentralization}
Adem~Efe Gencer, Soumya Basu, Ittay Eyal, Robbert Van~Renesse, and Emin~G{\"u}n
  Sirer.
\newblock Decentralization in bitcoin and ethereum networks.
\newblock In {\em International Conference on Financial Cryptography and Data
  Security}, pages 439--457. Springer, 2018.

\bibitem{6824541}
A.~{Gervais}, G.~O. {Karame}, V.~{Capkun}, and S.~{Capkun}.
\newblock Is bitcoin a decentralized currency?
\newblock {\em IEEE Security Privacy}, 12(3):54--60, 2014.

\bibitem{Karafiloski2017BlockchainSF}
Elena Karafiloski and Anastas Mishev.
\newblock Blockchain solutions for big data challenges: A literature review.
\newblock {\em IEEE EUROCON 2017 -17th International Conference on Smart
  Technologies}, pages 763--768, 2017.

\bibitem{kokoris2018omniledger}
Eleftherios Kokoris-Kogias et~al.
\newblock Omniledger: A secure, scale-out, decentralized ledger via sharding.
\newblock In {\em 2018 IEEE Symposium on Security and Privacy (SP)}, pages
  583--598. IEEE, 2018.

\bibitem{kwon2019impossibility}
Yujin Kwon, Jian Liu, Minjeong Kim, Dawn Song, and Yongdae Kim.
\newblock Impossibility of full decentralization in permissionless blockchains.
\newblock In {\em Proceedings of the 1st ACM Conference on Advances in
  Financial Technologies}, pages 110--123, 2019.

\bibitem{li2019incentivized}
Chao Li and Balaji Palanisamy.
\newblock Incentivized blockchain-based social media platforms: A case study of
  steemit.
\newblock In {\em Proceedings of the 10th ACM Conference on Web Science}, pages
  145--154, 2019.

\bibitem{Li2020ComparisonOD}
Chao Li and Balaji Palanisamy.
\newblock Comparison of decentralization in dpos and pow blockchains.
\newblock In {\em International Conference on Blockchain}, pages 18--32.
  Springer, 2020.

\bibitem{Miller2015DiscoveringB}
A.~Miller, James Litton, Andrew Pachulski, Neal Gupta, D.~Levin, N.~Spring, and
  B.~Bhattacharjee.
\newblock Discovering bitcoin’s public topology and influential nodes.
\newblock 2015.

\bibitem{nakamoto2019bitcoin}
Satoshi Nakamoto.
\newblock Bitcoin: A peer-to-peer electronic cash system.
\newblock Technical report, Manubot, 2019.

\bibitem{sagirlar2018hybrid}
Gokhan Sagirlar, Barbara Carminati, Elena Ferrari, John~D Sheehan, and Emanuele
  Ragnoli.
\newblock Hybrid-iot: Hybrid blockchain architecture for internet of things-pow
  sub-blockchains.
\newblock In {\em 2018 IEEE International Conference on Internet of Things
  (iThings) and IEEE Green Computing and Communications (GreenCom) and IEEE
  Cyber, Physical and Social Computing (CPSCom) and IEEE Smart Data
  (SmartData)}, pages 1007--1016. IEEE, 2018.

\bibitem{smith2017blockchain}
Tyler~D Smith.
\newblock The blockchain litmus test.
\newblock In {\em 2017 IEEE International Conference on Big Data (Big Data)},
  pages 2299--2308. IEEE, 2017.

\bibitem{srinivasan2017quantifying}
Balaji~S Srinivasan et~al.
\newblock Quantifying decentralization.
\newblock \url{
  https://news.earn.com/quantifying-decentalization-e39db233c28e}, [Accessed
  Nov. 2020], 2017.

\bibitem{tigani2014google}
Jordan Tigani and Siddartha Naidu.
\newblock {\em Google BigQuery Analytics}.
\newblock John Wiley \& Sons, 2014.

\bibitem{tschorsch2016bitcoin}
Florian Tschorsch and Bj{\"o}rn Scheuermann.
\newblock Bitcoin and beyond: A technical survey on decentralized digital
  currencies.
\newblock {\em IEEE Communications Surveys \& Tutorials}, 18(3):2084--2123,
  2016.

\bibitem{wang2019measurement}
Canhui Wang, Xiaowen Chu, and Qin Yang.
\newblock Measurement and analysis of the bitcoin networks: A view from mining
  pools.
\newblock {\em arXiv preprint arXiv:1902.07549}, 2019.

\bibitem{wu2019information}
Keke Wu, Bo~Peng, Hua Xie, and Zhen Huang.
\newblock An information entropy method to quantify the degrees of
  decentralization for blockchain systems.
\newblock In {\em 2019 IEEE 9th International Conference on Electronics
  Information and Emergency Communication (ICEIEC)}, pages 1--6. IEEE, 2019.

\bibitem{yue2016healthcare}
Xiao Yue, Huiju Wang, Dawei Jin, Mingqiang Li, and Wei Jiang.
\newblock Healthcare data gateways: found healthcare intelligence on blockchain
  with novel privacy risk control.
\newblock {\em Journal of medical systems}, 40(10):218, 2016.

\end{thebibliography}

\end{document}